\documentclass[seceq]{ptptex}

\def\beq#1{\begin{equation} \label{#1}}
\def\eeq{\end{equation}}

\def\bea{\begin{eqnarray}}
\def\eea{\end{eqnarray}}

\def\bra#1{\left\langle #1\right\vert}
\def\ket#1{\left\vert #1\right\rangle}

\def\MUU{ {\cal V } }
\def\MUD{ {\cal P } }
\def\eqref#1{(\ref{#1})}

\def\MM{{\cal M}}
\def\BB{{\cal B}}
\def\TT{\hbox{\small $\bar{\bf 3}{\bf 3}$}}
\def\SS{\hbox{\small $\bar{\bf 6}{\bf 6}$}}
\def\3s{\hbox{\small $\bar{\bf s}\bar{\bf 3}$}}
\def\6s{\hbox{\small $\bar{\bf s}{\bf 6}$}}
\def\ds{\displaystyle}

\title{
Exotic Hadrons in the Constituent Quark Model}

\author{
Harry J. \textsc{LIPKIN}%
}

\inst{
Department of Particle Physics \\
Weizmann Institute of Science, Rehovot 76100, Israel\\
 and\\
School of Physics and Astronomy \\
Raymond and Beverly Sackler Faculty of Exact Sciences \\
Tel Aviv University, Tel Aviv, Israel\\
and\\
High Energy Physics Division, Argonne National Laboratory \\
Argonne, IL 60439-4815, USA
}

\abst{

Exotic hadrons are important because their existence or absence can provide
important clues to understanding how QCD makes hadrons from quarks and gluons.
The first experimentally confirmed exotic will be the first hadron  containing
both $qq$ and $\bar q q$ pairs and the   first hadron containing color sextet
and color octet pairs. Theoretical models are not very useful because there is
no accepted model for multiquark systems with color-space correlations. The
constituent quark model is the only phenomenological model with predictive
power that has given experimentally tested universal  predictions for both
mesons and baryons. This paper reviews its explanation for why there are no 
bound exotics and its guidance to the search for heavy-flavored exotic
tetraquarks and pentaquarks. A possible supersymmetry between mesons and baryons
leading to meon-baryon mass relations not easily obtained otherwise is discussed.
}

\begin{document}

\maketitle
\section{Why exotics and the constituent quark model are important}

\subsection{Absence of exotics - an early clue to QCD}

The ``Goldhaber Gap" (no $K^+N$ resonances) led Gell-Mann to quarks. The
experimental hadron spectrum today still shows no exotic bound states with exotic quantum
numbers. The only bound states are
color singlet $(q\bar q)$ mesons and baryons containing color
antitriplet $(q q)$ pairs. Finding the first exotic would help our understanding how QCD makes
hadrons from quarks and gluons. It would be the first multiquark hadron
containing  both $q$ and $\bar q$ and the first hadron containing both color
sextet $qq$ pairs and color octet $(\bar q q)$ pairs. 

Some crucial questions
\begin{enumerate} 
\item How does QCD make hadrons from quarks and gluons? 
\item DOES QCD make hadrons from quarks and gluons?
\item Do we need more than the standard model?
\item  Do we need another symmetry or supersymmetry?
\item  Why are there no bound exotics?
\item How can a pion be both a Goldstone Boson and 2/3 of a nucleon?
\end{enumerate}

\subsection{The constituent quark model helps understanding how QCD makes
hadrons}

Experimental data show us mesons and baryons made of same quarks with
flavor-independent interactions for five flavors.  The constituent quark model
is the only hadron model with demonstrated reliable predictive power,  includes
five-flavor meson-baryon universality and can treat states containing both $qq$
and $\bar q q$. The model makes testable experimental predictions for the
two-body interactions $V(q\bar  q)_8$ and $V(q q)_6$ and for states containing
both  $qq$ and $\bar q q$. No other model makes such predictions where no
experimental data are yet available and which are tested by the  presence  or
absence of exotics. 

\section{What are constituent quarks - A BCS Approach 
from John Bardeen} 

Quarks are quasiparticle degrees of freedom describing  low-lying elementary
excitations of hadronic matter.  BCS theory was not gauge invariant - so what?
Bardeen knew it had the right physics. Anderson explained the
broken gauge invariance - found the Higgs mechanism.   Particle physicists
don't recognize that Anderson found Higgs. In 1960 Bardeen noted Nambu's
interesting application of superconductivity ideas to particle physics.  But
at the 1960 International High energy (Rochester) Conference   particle
physicists showed no interest in Nambu's work on symmetry breaking.  BCS theory
must be obtainable from the QED Lagrangian, but nobody knew how to do it.
Summing Feynman diagrams or putting electrons on a lattice still cannot lead to
BCS. 

The constituent quark model is not relativistic - so what?
The experimental data show it has the right physics.
Getting constituent quarks from QCD or the lattice might
be as difficult as getting BCS from QED.

That meson and baryon masses are related because they are made of the same
quarks was first pointed out by Sakharov and Zeldovich\cite{SakhZel} 
and completely  independently
rediscovered\cite{hjl78,einstein}.
They noted 
that $\Lambda$ and $\Sigma$ were made of same quarks,
$anticipated$ QCD
and explained their mass difference with a flavor-dependent hyperfine 
interaction. Their constituent quark mass formula
gave two surprising\cite{SakhZel,hjl78,einstein} meson - baryon mass 
relations which were later followed by
a number of additional successful quark model  relations
\cite{NewPenta,szmassqcd}  with no simple explanations from QCD. 
\beq{szhyp}
M = \sum_i m_i + \sum_{i>j}  {{\vec{\sigma}_i\cdot\vec{\sigma}_j}\over{m_i\cdot
m_j}}\cdot v^{hyp}_{ij}
\end{equation}

These quasiparticle excitation of the QCD vacuum
carry
the same spin and flavor quantum numbers as current quarks.
Mesons are $q \bar q$ pairs, baryons are $3q$ and nothing else.
Their effective quark masses include all interaction energies 
except the color hyperfine energy and have
the same values for mesons and baryons.
Their spin-dependent interactions are given by effective moments equal to
Dirac moments with the same effective quark mass.
Hadron magnetic moments are the vector sums of quark moments.
Hyperfine splittings are proportional to products of the same (color) moments.
 
The experimental successes of model - challenges for QCD -
include three magnetic moment predictions with no free parameters\cite{NewPenta}
\beq{barmom1}
-1.46 =
{\mu_p \over \mu_n} =
-{3 \over 2}; ~ ~ ~
-0.61
\,{\rm n.m.}=\mu_\Lambda=
-{\mu_p\over 3} {{M_{\Sigma^*} - M_\Sigma} \over{M_\Delta - M_N}}
=-0.61 \,{\rm n.m.}
\end{equation}
\beq{barmom3}
\mu_p+\mu_n= 0.88 \,{\rm n.m.}
={2M_{\scriptstyle p}\over M_N+M_\Delta}=0.865 \,{\rm n.m.}
\end{equation}

The last prediction sets absolute scale of magnetic moments.

QCD calculations have not yet explained such remarkably successful simple 
constituent quark model results.  It is not a nonrelativistic nor potential
quark model.  The internal structures and space-time properties of constituent
quarks are not defined. What they really are remain as challenges for QCD.

A completely different experimental confirmation of this picture is seen in the
successful relations between meson and baryon total cross
sections\cite{sigtot,nuhyp}; e.g. the predictions for
total cross sections 
at $P_{lab}$ = 100 GeV/c,
\beq{sigtotp}
38.5 \pm 0.04 \,{\rm mb} = \sigma_{tot}(pp) =
3 \sigma_{tot}(\pi^+p) -{3\over 2}\sigma_{tot}(K^-p) =39.3 \pm 0.2 \,{\rm mb}
\end{equation}
\beq{sigtotsig}
33.1 \pm 0.31 \,{\rm mb} = \sigma_{tot}(\Sigma p) =
{3\over 2}\{\sigma_{tot}(K^+p) + \sigma_{tot}(\pi^-p) -\sigma_{tot}(K^-p)\}
=33.6 \pm 0.16 \,{\rm mb}
\end{equation}
\beq{sigtotxi}
29.2 \pm 0.29 \,{\rm mb} = \sigma_{tot}(\Xi p) =
{3\over 2}\sigma_{tot}(K^+p) =28.4 \pm 0.1 \,{\rm mb}
\end{equation}
But we still do not know what the constituent
quark is.

\section{The LS transformation -  A new meson-baryon supersymmetry?}
\subsection{The prediction for the newly discovered  $\Sigma_b$ baryons}

A new challenge demanding explanation from QCD is posed by 
the remarkable agreement between the experimental masses\cite{CDF_Sigma_b} 
of the newly discovered  $\Sigma_b^+$ and $\Sigma_b^-$ 
and the 
prediction\cite{NewPenta,sigmab} from meson masses, 
\beq{newpred2}
\begin{array}{ccccccc}
\displaystyle
{{M_{\Sigma_b} - M_{\Lambda_b}}\over{(M_\rho - M_\pi)-(M_{B^*}-M_B)}} &=&  
\displaystyle
{{M_{\Sigma_c}{-}M_{\Lambda_c}}\over{(M_\rho {-} M_\pi){-}(M_{D^*}{-}M_D)}}
&=&
\displaystyle
{{M_{\Sigma}{-}M_\Lambda}\over{(M_\rho {-} M_\pi){-}(M_{K^*}{-}M_K)}}
\\
0.32
&\approx &
0.33
&\approx& 
0.325
\end{array}
\end{equation}

New successful relations, (\ref{newpred2}) and others described below indicate
some light antiquark-diquark supersymmetry\cite{sigmab} between meson and
baryon states not simply described by QCD treatments, which  treat meson and
baryon structures very differently. 
This liqht quark supersymmetry transformation, denoted here by $T^S_{LS}$,
connects a meson denoted by $\MM(\bar q Q_i)$ and a baryon denoted by 
$B([qq]_S Q_i)$ both containing the same valence quark of some fixed flavor
$Q_i$, $i=(u,s,c,b)$ and  a light color-antitriplet ``brown muck" state with 
the flavor and baryon quantum numbers respectively of an antiquark $\bar  q$
($u$ or $d$) and two light quarks coupled to a diquark of spin $S$.  No model
is assumed for the valence quark nor for the brown muck antitriplet which is
coupled to it. This goes beyond the simple constituent quark model and holds
also for the quark-parton model in which the valence is carried by a current
quark and the rest of the hadron is a complicated mixture of quarks and
antiquarks.

\beq{numesbar}
T^S_{LS} \cdot \MM(\bar q Q_i) \quad \equiv \quad \BB([qq]_S Q_i)
\end{equation}
The mass difference between the meson and baryon related by this $T^S_{LS}$
transformation has been shown\cite{sigmab} to be independent of the quark
flavor $i$ for all four flavors $(u,s,c,b)$ for the two cases of
spin-zero\cite{szmassqcd} $S=0$ and  spin-one\cite{sigmab} $S=1$ diquarks, 
when we use weighted averages of hadron masseswhich cancel their hyperfine
contributions,
 
\beq{tildev}
\tilde M(V_i)\equiv {{3M_{\MUU_i} + M_{\MUD_i} }\over {4}}; 
\qquad
\tilde M(\Sigma_i)\equiv {{2M_{\Sigma^*_i} + M_{\Sigma_i}}\over {3}} ; 
\qquad
\tilde M(\Delta)\equiv{{2 M_\Delta +M_N}\over {3}}
\end{equation}
   
\beq{mesbardif}
\begin{array}{ccccccc}
M(N) - \tilde M(\rho) &=& M(\Lambda) - \tilde M(K^*) &=&
M(\Lambda_c) -  \tilde M(D^*) &= &
M(\Lambda_b) - \tilde M(B^*)
\\
323~\rm{MeV} &\approx& 321~\rm{MeV} &\approx&312~\rm{MeV}
&\approx& 310~\rm{MeV}
\end{array}
\end{equation}
\beq{mesbardif2}
\begin{array}{ccccccc}
\tilde M(\Delta) - \tilde M(\rho) &=& M(\Sigma) - \tilde M(K^*) &=&
\tilde M(\Sigma_c) -  \tilde M(D^*) &= &
\tilde M(\Sigma_b) - \tilde M(B^*)
\\
517.56 ~\rm{MeV} &\approx& ~526.43\rm{MeV} &\approx& 523.95  ~\rm{MeV}
&\approx& 512.45 ~\rm{MeV}
\end{array}
\end{equation}

The ratio of the hyperfine splittings of mesons and baryons 
related by $T^1_{LS}$ is also independent of the quark\cite{szmassqcd,boaz}
flavor $i$ for all four flavors $(u,s,c,b)$,

\beq{LShyperfine}
\begin{array}{ccccccc}
\displaystyle
\frac{M_\rho - M_\pi}{M_\Delta - M_N}&= &
\displaystyle
\frac{M_{K^*}-M_K}{M_{\Sigma^*} - M_\Sigma}&= &
\displaystyle
\frac{M_{D^*}-M_D}{M_{\Sigma_c^*} - M_{\Sigma_c}}&= &
\displaystyle
\frac{M_{B^*}-M_B}{M_{\Sigma^*_b}-M_{\Sigma_b}}
\\
2.17\pm0.01 &=& 2.08\pm0.01  &= & 2.18 \pm 0.01 &= & 2.15 \pm 0.20
\end{array}
\end{equation}

That masses of boson and fermion states containing quarks of four different
flavors, $u,d,s,b$, related by this transformation (\ref{numesbar}) satisfy 
simple relations like (\ref{newpred2}), (\ref{mesbardif}), (\ref{mesbardif2})
and (\ref{LShyperfine}) with no free parameters can hardly be accidental. They
remain a challenge for QCD perhaps indicating some boson-fermion or
antiquark-diquark supersymmetry. Any model for hadron spectroscopy which treats
mesons and baryons differently or does not yield agreement with data for  all
five flavors is missing essential physics.   

\section {The Nambu color exchange interaction (1966) anticipating QCD}  

\subsection{Nambu's Theorem (1966) : No exotics. 
Only lowest color singlets are stable} 

Nambu\cite{Nambu} considered colored quarks interacting via a non-abelian SU(3)
gauge field in the theory now called QCD. 
The color-exchange interaction (one gluon exchange) between two constituents 
$i$ and $j$ is for a multiquark system,
\beq{nambumesbar}
 V_{cx}(r_{ij}) = {V\over2} \cdot\sum_{i \not= j}
\vec \lambda_c^i\cdot\vec \lambda_c^j \cdot v(r_{ij})
\end{equation}    
where $i$ and $j$ can be either quarks or antiquarks, $ v(r_{ij})$  depends on
the space
and spin variables of the
constituents but is the same for all pairs, independent of $i$ and
$j$, $\vec \lambda_c^i$ is the color SU(3) generator and 
$\vec \lambda_c^i\cdot\vec \lambda_c^j$ denotes the
scalar product in color space.
In lowest order neglecting color-space and color-spin 
correlations 
\beq{nambutot}
  \frac{\langle V_{cx}(tot)\rangle}{\langle v(r) \rangle} 
={V\over2} \cdot  \left[ \sum_{i j}
\vec \lambda_c^i\cdot\vec \lambda_c^j 
-\sum_{i}
(\lambda_c^i)^2\right]
={V\over2}
\cdot \left[\boldmath (\lambda_C)^2 -\sum_{i}\boldmath
(\lambda_c^i)^2\right]
\end{equation}
where $\lambda_C$
is the generator of the color SU(3) group for the whole multiquark system.
The lowest states having N constituents are color singlet states having
$\lambda_C = 0$. All states with the same overall color have the same potential
energy. 

Nambu's theorem predicts that the lowest bound hadrons are color singlet 
$(q\bar q)_1$ mesons  and color singlet baryons in which each quark pair is
coupled to a color antitriplet $(q q)_{\bar 3}$.  Other color singlets are not
bound as they can gain kinetic energy by breaking up into color singlets. 

\subsection{Color-space correlations and tetraquarks with the Nambu interaction} 

The tetraquarks are the simplest multiquark color singlet system.
They exhibit features missed in other models suggested for these states and
provide useful models for mesons containing heavy quarks; e.g. color sextet
quark-quark couplings.

The first treatment of states containing both $qq$ and $\bar q q$ found new
forces between color singlet hadrons in a tetraquark\cite{triex}  $qq\bar q
\bar q$ with the Nambu interaction (\ref{nambumesbar}). There are two color
singlet couplings for  $qq\bar q \bar q$ system. The $qq$ can be coupled either
to a color $\bar 3$ or color $\bar 6$ and combined respectively with the $\bar
q \bar q$ in color $3$ or color $\bar 6$ to make a color singlet.  The Nambu
interaction gives the same energy for all color couplings in systems with
maximum space symmetry.  

The color sextet diquark is not present in normal baryons. The Nambu
interaction gives a repulsive $qq$ interaction exactly compensated by  $q \bar
q$ attraction in the $6 \bar 6$ configuration. Potential energy can be gained
in $6 \bar 6$ by breaking space symmetry and keeping $qq$ spacing  larger that
$q \bar q$. The model\cite{triex}  showed this gain was not sufficient to
overcome kinetic energy with equal quark masses and reasonable potentials. 

Explicit calculations using a harmonic oscillator  spatial  dependence of the
operator $V$  in the interaction (\ref{nambutot})with broken space
symmetry\cite{newpentel} show that the $(\SS)$ state is always considerably
below the  $(\TT)$ state. For sufficiently unequal quark masses the $(\SS)$ is
also below the two-meson threshold. 

Calculations for the $(u s \bar d \bar c)$ system with four different flavors
and four different masses show how the difference between the tetraquark mass
and the  mass of two separated mesons depends upon the quark masses.

When all the spatial separations are equal, $r_{us}^2 = r_{\bar c \bar d}^2 =
r_{u \bar c}^2= r_{s \bar c}^2 = r_{u \bar d}^2= r_{s \bar d}^2$ the $q\bar q$
interaction in the state $\SS$ is 25\% stronger than the $q\bar q$ interaction
in the separated  two-meson state $2M$. This additional attraction is balanced
exactly by the $us$ and $\bar c \bar d$ repulsions as required by Nambu's
theorem. But more additional attraction  is obtainable by breaking space
symmetry and making the mean $us$ and $\bar c \bar d$ distances larger than the
mean $q\bar q$ distance. 

The ratios of ground state energies for the $\TT$ and $\SS$ systems 
to the energy of the two meson state are obtained by
setting $m_u = m_d$ and substitute 
constituent quark masses obtained by
fitting ground state meson and baryon spectra\cite{NewPenta}
\beq{qmasses}
m_u =  360
\hbox{\ MeV};
m_s=  540
\hbox{\ MeV};
m_c= 1710
\hbox{\ MeV}
;
m_b=  5050
\hbox{\ MeV}
\end{equation}
\beq{usdcrat}
\displaystyle
{{E_g(\TT\,usdc)}\over{E_g(2M\,usdc)}} 
=1.18 ; ~ ~ ~
\displaystyle
{{E_g(\SS\,usdc)}\over{E_g(2Musdc)}} 
=1.11
\end{equation}

The $(\SS)$ state is considerably below the $(\TT)$ state 

When heavier quark states are included the ratio of the $(\SS)$ mass to the
two-meson threshold is seen to drop with increasing quark mass and is actually 
below the threshold for the $cubu$ and $bubu$ tetraquarks.
\beq{tetrarat}
\displaystyle
{{E_g(\SS\,subu)}\over{E_g(2M\,subu)}} =
1.077
; ~ ~ ~
{{E_g(\SS\,cucu)}\over{E_g(2Mcucu)}} 
=1.036
; ~ ~ ~
\displaystyle
{{E_g(\SS\,cubu)}\over{E_g(2M\,cubu)}} 
= 0.975
; ~ ~ ~
{{E_g(\SS\,bubu)}\over{E_g(2M\,bubu)}} 
= 0.891
\end{equation}

In this approximation the mass of the $c u \bar c \bar u$ tetraquark  is only
4\% above the the $D\bar D$ threshold and  the $b u \bar b \bar u$ tetraquark 
is well below the $B\bar B$ threshold. Such low-mass thresholds may well be
found in a more exact calculation including spin effects. That they might be
found in the experimental spectrum must be taken seriously.
The color-space correlation contributions to the energy may well be more
important than  the color-magnetic energy commonly used in model calculations.

This casts doubt on tetraquark calculations
neglecting color space correlations; e.g. those for the X(3872) resonance.  
QCD tells us space symmetry is broken by the strongly attractive  $q\bar q_1$ 
interaction, much stronger than the attractive $qq_{\bar 3}$ interaction.

\subsection{Color-spin correlations with 
Nambu's interaction}

The DeRujula-Georgi-Glashow Model\cite{DGG} introduced a color-spin hyperfine 
interaction and obtained remarkable agreement with nonexotic hadron spectra,
including masses, spin splittings and magnetic moments.
Jaffe\cite{Jaffe} extended the DGG model \cite{DGG} to treat multiquark systems
using the Nambu interaction (\ref{nambumesbar}) to
predict $V(q\bar q)_8/V(q\bar q)_1$
and  $V(q q)_6/V(qq)_{3*}$ which are not obtainable from the nonexotic hadron
spectrum.
The Jaffe model explains striking features of the hadron spectrum.
Pauli principle ``flavor antisymmetry" requires repulsive short-range
color-magnetic interaction between same-flavor quark pairs 
thus explaining the absence of low-lying bound exotics
like a dipion with a mass less than two pion masses or a dibaryon
bound by 100 MeV. 
Only the
short-range color-magnetic interaction can produce
multiquark binding. Best candidates with a minimum number of
repulsive same-flavor pairs are Jaffe's  $H$ $(uuddss)$
dibaryon and the anticharmed strange
$(uuds\bar c)$ pentaquark. The $\Theta^+$ pentaquark $(uudd \bar s)$ has too 
many same-flavor quark pairs and was not considered.

\section{Guide to searches for exotics}

\subsection{The $\Theta^+ (u u d d \bar s)$ pentaquark} 

The  $\Theta^+$ is discussed extensively elsewhere\cite{jenmalt}. A color-space correlation
similar to those for our tetraquarks is found in  one calculation\cite{hosaka}
using a harmonic oscillator hamiltonian with an additional spin-dependent
interaction in a model space of 15,000 basic single-particle shell-model wave
functions. The mean quark-antiquark distance is much smaller than the mean
quark-quark distance. The one-body r.m.s. radius measured from the center of
mass is 1.10 Fm. for $u$,$d$ quarks and 0.72 Fm for the $\bar s$ antiquark
while the corresponding radius is 0.69 Fm for the $(0s)^5$ configuration. 

That the one-body r.m.s. radius of the multiquark system is larger than
expected from normal hadrons implies a short-range color-magnetic
interaction weaker than in normal hadrons. The conventional practice
of using spin splittings from normal hadrons to normalize the color-magnetic
interaction\cite{NewPenta,Jaffe} is therefore questionable.

\subsection{How to search for stable exotic baryons} 

Ashery at E791 searched for the $\bar c uuds$ pentaquark\cite{E791Col},
found events, but not enough to be convincing. 
The possible existence of this pentaquark is still open.
Better searches are needed. 
A simple clean search for protons from a secondary vertex would see the 
signature for weakly decaying baryon and would immediately find all new weakly 
decaying baryons. 
Jaffe model calculations 
found no reason to look for the $\Theta^+ (uudd \bar s)$ pentaquark.

\subsection{Experimental candidates for tetraquark searches}

The best candidates for experimental detection are the $c u
\bar b \bar u$ and $s u \bar b \bar u$ tetraquarks which cannot
decay strongly or electromagnetically into light quark mesons.

     A $b q \bar c \bar q$ tetraquark with isospin 1 and a mass below the $B
\bar D$ threshold but above the mass of the $B_c \pi$ system can decay strongly
into a $B_c \pi$ with a width limited by phase space. Both isospin
states below the $B_c \pi$ threshold should be narrow and
decay electromagnetically into $B_c \gamma$. A spin-zero state
must decay by $e^+e^-$ emission. An $I=0$ state above  $B_c \pi$
threshold can  decay  into $B_c \pi$ via isospin violation.  These
considerations apply also to states below the $B^* \bar D$ threshold that
cannot decay into $B\bar D$. 
A $b q\bar c \bar q$ tetraquark below the $B_c$ mass can decay only weakly and
can  appear in invariant mass
plots of final states like $J/\psi e\nu$ as an additional mass peak along with
the $B_c$. Isovector tetraquarks like $b u \bar c \bar d$ or $b d \bar c \bar
u$ have exotic final states with wrong charges, like $J/\psi \eta$ or $J/\psi
\pi^- \pi^-$.
Searches for such tetraquarks are of interest in
experiments observing the $B_c$.
One can look for monoenergetic photons or pions
emitted together with a $B_c$, a doublet structure of the $B_c$ mass and
exotic $B_c$ decays.
Analogous considerations hold for $b q \bar s \bar q$ tetraquarks with 
masses below $B K$ threshold.

    Exotic multiquark states with color-space correlations have a larger
extension in space than normal hadrons, Production of such states make be
difficult as they can be easily broken  up by final state interactions or
rescattering.

\section{Conclusions}

The Nambu interaction 
explains the absence of strongly bound exotics and provides guide lines for
future searches.
Color-space correlations may
be more important than color-flavor-spin correlations dominating
other treatments.
Basic QCD physics implies the $q\bar q$
interaction observed in mesons is much stronger in multiquark systems than the $q q$ interaction
observed in baryons, produces  admixtures of quark states absent
in normal baryons\cite{NewPenta,jenmalt} and destroys all diquark
structures.
Color-space correlated tetraquarks may be found in
mesons containing heavy quarks.

Searches for tetraquarks 
in the $B_c$ system may find $b q \bar c
\bar q$ tetraquarks below the $B\bar D$ threshold. Possible exotic
signatures include strong or electromagnetic decays into a $\pi B_c$  or
$\gamma B_c$, weak decays producing additional peaks in the mass spectrum of
$B_c$ decay final states or final states with exotic electric charge, like
$J/\psi \eta$ or $J/\psi \pi^{-} \pi^{-}$.

\section*{Acknowledgements}
It is a pleasure to acknowledge discussions with Atsushi Hosaka, Marek Karliner,
Boaz Keren-Zur and Takashi Nakano.

%


\begin{thebibliography}{99}


\bibitem{SakhZel}{
Ya.B. Zeldovich and A.D. Sakharov, Yad. Fiz {\bf 4}(1966)395; 
Sov. J. Nucl. Phys. {\bf 4}(1967)283.}

\bibitem {hjl78} Harry J. Lipkin, 
Phys. Rev. Lett. {\bf 41}, 1629 (1978).


\bibitem {einstein} Harry J. Lipkin, 
in To fulfill a vision: Jerusalem Einstein Centennial
Symposium on Gauge Theories and Unification of Physical Forces,
edited by Yuval Ne'eman, (Addison Wesley Publishing Company, 1980) p. 139

\bibitem{NewPenta}
M. Karliner and H.J. Lipkin, hep-ph/0307243; condensed version in 
Phys.\ Lett.\ B {\bf 575} (2003) 249.


\bibitem {szmassqcd}
M. Karliner and H.J. Lipkin, hep-ph/0608004

\bibitem{sigtot} {Harry J. Lipkin, Phys.Lett.B242:115,1990}

\bibitem{nuhyp} {Harry J. Lipkin,
hep-ph/9911259
In Hyperon 99, Proceedings of the Hyperon Physics Symposium Hyperon 99,
Fermilab, Batavia, Illinois, September 27-29 (1999( edited by  D. A. Jensen
and E.Monnier, FERMILAB-Coonf-00/059-E (2000) 87}


\bibitem{CDF_Sigma_b}
{\tt http://www-cdf.fnal.gov/physics/new/bottom/060921.blessed-sigmab/}


\bibitem{sigmab}Marek Karliner and Harry J. Lipkin, hep-ph/0611306 

\bibitem {boaz} Boaz Keren-Zur, hep-ph/0703011

\bibitem{Nambu} Y. Nambu, 
in Preludes in Theoretical Physics, edited by
A. de Shalit, H. Feshbach and L. Van Hove, (North-Holland Publishing Company,
Amsterdam, 1966), p. 133
\bibitem{triex} H.J. Lipkin, \PLB{45,1973,267}
\bibitem{newpentel}
 Marek Karliner and Harry J. Lipkin.Phys.Lett. B638 (2006) 221, hep-ph/0601193




\bibitem{DGG}{A. De Rujula, H. Georgi and S.L. Glashow, Phys. Rev. D12
(1975) 147}
\bibitem{Jaffe}{R. L. Jaffe, {\it Phys. Rev. Lett.} {\bf 38},  195 (1977)}
\bibitem{jenmalt}
For a review of the considerable theoretical literature on
pentaquark models and an in-depth discussion, see Byron K.
Jennings and Kim Maltman, hep-ph/0308286
\bibitem{hosaka}E. Hiyama, M. Kamimura, A. Hosaka, H. Toki and M. Yahiro,
hep-ph/0507105
\bibitem{E791Col}{E.M. Aitala et al.,
E791 Collaboration,
FERMILAB-Pub-97/118-E, {\it Phys. Lett.}  {\bf B448}, 303  (1996). }

\end{thebibliography}
\end{document}